\documentclass[sigconf]{acmart}
\usepackage{wrapfig}
\usepackage{multirow}
\usepackage{amsmath,graphicx}

\AtBeginDocument{%
  }

\copyrightyear{2023}
\acmYear{2023}
\setcopyright{acmlicensed}
\acmConference[ICCIP 2023]{2023 the 9th International Conference on Communication and Information Processing (ICCIP)}{December 14--16, 2023}{Lingshui, China}
\acmBooktitle{2023 the 9th International Conference on Communication and Information Processing (ICCIP) (ICCIP 2023), December 14--16, 2023, Lingshui, China}
\acmDOI{10.1145/3638884.3638917}
\acmISBN{979-8-4007-0890-9/23/12}

\begin{document}

\title{Adversarial Data Augmentation for Robust Speaker Verification}

\author{Zhenyu Zhou}
\affiliation{%
  \institution{Beijing University of Post Telecommunications}
  \country{China}}
\email{buptzzy2022@bupt.edu.cn}

\author{Junhui Chen}
\affiliation{%
  \institution{Beijing University of Post Telecommunications}
  \country{China}}
\email{cjharryyyds@bupt.edu.cn}

\author{Namin Wang}
\affiliation{%
  \institution{Huawei Cloud}
  \country{China}}
\email{wangnamin@huawei.com}

\author{Lantian Li}
\affiliation{%
  \institution{Beijing University of Post Telecommunications}
  \country{China}}
\email{lilt@bupt.edu.cn}

\author{Dong Wang}
\affiliation{%
 \institution{CSLT@BNRist, Tsinghua University}
 \country{China}}
\email{wangdong99@mails.tsinghua.edu.cn}


\begin{abstract}

Data augmentation (DA) has gained widespread popularity in deep speaker models due to its ease of implementation and significant effectiveness.
It enriches training data by simulating real-life acoustic variations, enabling deep neural networks to learn speaker-related representations 
while disregarding irrelevant acoustic variations, thereby improving robustness and generalization.
However, a potential issue with the vanilla DA is \emph{augmentation residual},
i.e., unwanted distortion caused by different types of augmentation.

To address this problem, this paper proposes a novel approach called adversarial data augmentation (A-DA) which combines DA with adversarial learning.
Specifically, it involves an additional augmentation classifier to categorize various augmentation types used in data augmentation.
This adversarial learning empowers the network to generate speaker embeddings that can deceive the augmentation classifier,
making the learned speaker embeddings more robust in the face of augmentation variations.
Experiments conducted on VoxCeleb and CN-Celeb datasets demonstrate that our proposed A-DA outperforms standard DA in both augmentation matched and mismatched test conditions, 
showcasing its superior robustness and generalization against acoustic variations.

\end{abstract}



\maketitle

\section{Introduction}

Automatic speaker verification (ASV) is aimed at verifying the claimed identity of a speech segment~\cite{campbell1997speaker,hansen2015speaker}. 
Over decades of research, current ASV systems have made significant strides, 
primarily owing to the continuous accumulation of speech data and the prevalence of speaker embedding models based on deep neural networks (DNNs)~\cite{reynolds2002overview,kinnunen2010overview,bai2021speaker}. 
The x-vector architecture and its variants are among the most widely adopted deep embedding models~\cite{snyder2018x,garcia2019x}. 
Recently, with carefully designed architectures and training techniques, 
deep embedding models have achieved state-of-the-art performance in numerous ASV evaluation tasks~\cite{sadjadi20222021,huh2023voxsrc}.

Despite these significant advancements, current ASV systems still encounter numerous challenges in terms of robustness when deployed in real-world applications.
One major challenge involves the intricate interplay of speaker traits with complex and diverse acoustic variations, 
including background noise, music, multi-speaker conversations, and more.
These acoustic variations result in unpredictable shifts in speaker embedding models, leading to performance degradation.

To tackle this challenge, researchers have introduced a range of methods.
One of the most successful and widely used techniques is data augmentation (DA)~\cite{snyder2018x,desplanques2020ecapa}, 
mainly due to its ease of implementation and significant effectiveness.
The purpose of DA is to enrich the quantity and diversity of the training data by simulating complex acoustic variations.
Current DA methods can generally be categorized into two groups.
One group involves operations on the raw speech signal, such as adding additive noise and reverberation~\cite{amini2021data,ko2017study}, 
speed perturbation~\cite{yamamoto19speaker,chen2022build}, volume perturbation~\cite{huang2019exploring}, and more.
The other group augments the spectrogram by applying random masks in the time and frequency domains~\cite{park2019specaugment,wang2020investigation}.

All of these DA methods have been demonstrated effective, especially in DNN-based speaker embedding models.
With a large amount of augmented training data and guided by the training objective of maximizing the discrimination between different speakers, 
DNNs can comprehensively learn speaker-related representations while disregarding irrelevant acoustic variations~\cite{li2017deep}. 
This, in turn, enhances the robustness and generalization of speaker embedding models across various acoustic conditions.

However, despite the success of DA in enhancing the robustness of speaker verification, 
current DA methods suffer from a potential drawback known as \emph{augmentation residual}. 
This means that when training with data augmented under a specific augmentation type, the speaker embeddings might be systematically distorted. This is because the speaker discrimination loss, e.g., cross-entropy, does not impose any invariance constraint on the embeddings, and a low cross entropy could be still obtained even with the distorted embeddings. However, the unwanted distortion may lead to reduced generalizability. 

To address the issue of augmentation residual, this paper proposes a novel training strategy called \emph{adversarial data augmentation (A-DA)}.
This approach draws inspiration from the success of domain adversarial training in speaker recognition tasks, such as 
unsupervised domain adaptation~\cite{ganin2015unsupervised,wang2018unsupervised,huh2020augmentation} 
and domain-invariant representation learning~\cite{wang2021adversarial,zhang2022cross}.
Generally, these domain adversarial training methods employ a gradient reversal layer to remove the domain variation 
and project different domain data into the same subspace, resulting in domain-invariant and speaker-purified representations.

In this paper, we combine data augmentation with adversarial training to improve the robustness of deep speaker models in the presence of acoustic variations.
Initially, the standard DA is applied to diversify the training data fed into DNNs. 
Subsequently, two objective losses are integrated. 
One is the speaker classification loss with cross-entropy, used to distinguish between different speakers in the training data.
The other is the augmentation classification loss with binary cross-entropy, used to categorize different acoustic types applied during data augmentation.
Then a gradient reversal layer is involved in the back-propagation process of the acoustic classification loss.
This process of adversarial learning allows speaker embeddings generated by the network encoder, 
which are augmented with different types of acoustic conditions, 
to deceive the acoustic classifier.
In other words, this enhances the capability of the network to learn speaker embeddings that remain robust in the face of acoustic variations.

Our experiments are firstly conducted on the VoxCeleb dataset~\cite{chung2018voxceleb2} 
and utilized noise, speech, and music from the MUSAN dataset~\cite{snyder2015musan} for data augmentation. 
The results demonstrate that under the training-matched augmentation conditions, 
our proposed A-DA method is more robust compared to the standard DA method.
Furthermore, we use cafe and car noises from the THCHS-30 dataset~\cite{wang2015thchs} and also the official CN-Celeb evaluation set~\cite{fan2020cn}
to validate the generalizability of the A-DA method against unseen augmentation test conditions.
Experimental results consistently show a performance advantage for our proposed A-DA method in these unseen augmentation test conditions, 
highlighting its superior generalization against acoustic variations.

\section{Related work}

In the field of speaker verification, researchers have explored various methods of both data augmentation and adversarial training to enhance model robustness.

Regarding data augmentation (DA), researchers perform a series of manipulations on the raw data, such as adding additive perturbation and making random disruptions
~\cite{amini2021data,ko2017study,yamamoto19speaker,chen2022build,huang2019exploring,park2019specaugment,wang2020investigation}. 
These manipulations are designed to simulate complex acoustic variations, effectively increasing the volume and diversity of training data. 
By leveraging the powerful feature learning capability of deep neural networks, deep embedding models can learn speaker traits that are insensitive to acoustic variations, thereby improving robustness in the presence of complex acoustic conditions.

Adversarial training (AT) was initially applied in unsupervised domain adaptation tasks in speaker verification to 
address the distribution mismatch issue between the source and target domain~\cite{wang2018unsupervised}.
The training objective involves minimizing speaker classification loss in the source domain 
while maximizing the domain classification loss between the source and target domains. 
This approach has been further extended to multi-domain speaker verification, 
aiming to learn domain-agnostic speaker representations that enhance robustness across multiple domains.

However, the combination of DA and AT in the field of speaker verification is a relatively under-explored research area.
Recently, Jaesung et al.~\cite{huh2020augmentation} proposed an augmentation adversarial training method for self-supervised speaker recognition. 
In this method, which assumes without speaker labels, 
data augmentation and negative sampling are used for contrastive learning to extract speaker discriminative representations. 
Additionally, adversarial training is employed to explicitly guide the network in learning speaker representations that are insensitive to augmentation, 
making the learned speaker embeddings more robust.

The core idea of this paper is similar to Jaesung et al.~\cite{huh2020augmentation} but differs in two key aspects.
Firstly, this paper focuses on the supervised speaker verification framework rather than self-supervised, 
providing a more direct insight into the interplay between DA and AT. 
Secondly, this paper goes beyond demonstrating the effectiveness of the combination of DA and AT under seen augmentation conditions; 
it also validates the generalization of their combination in unseen augmentation conditions.

\section{Adversarial Data Augmentation}

\begin{figure*}[h]
  \centering
  \includegraphics[width=\linewidth]{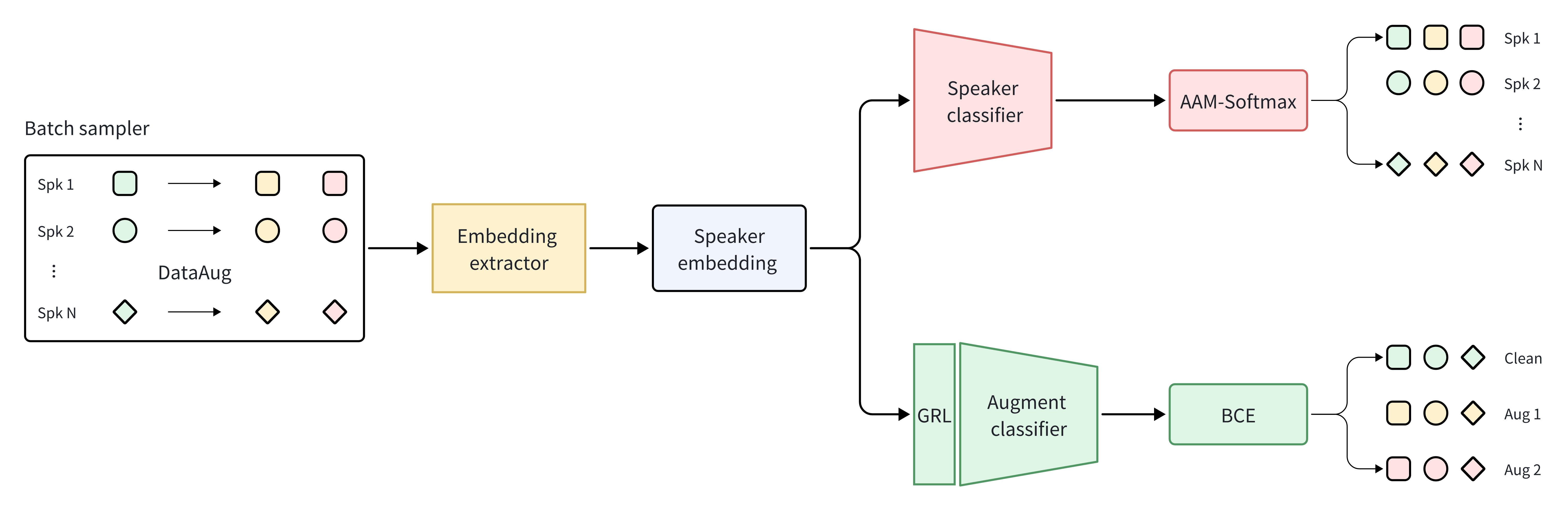}
  \caption{Illustration on the training strategy of our proposed adversarial data augmentation method.}
  \label{fig:ada}
\end{figure*}

This section describes the proposed adversarial data augmentation method, as illustrated in Figure~\ref{fig:ada}.
We first introduce the data augmentation module, followed by the presentation of the batch sampler strategy for training. 
Subsequently, we describe the adversarial training on augmented data, which leverages an augmentation classifier and a gradient reversal layer 
in addition to the speaker embedding extractor. Finally, the entire neural network is trained with the dual objective 
of minimizing the speaker classification loss and maximizing the augmentation classification loss.

\subsection{Data augmentation}

In this study, we implement DA using additive noises. 
Specifically, we use three types of noises (noise, music, and speech) from the MUSAN dataset~\cite{snyder2015musan} and 
and follow the augmentation process as outlined in~\cite{snyder2018x}.
More details please refer to ~\footnote{https://github.com/kaldi-asr/kaldi/tree/master/egs/voxceleb} and~\footnote{https://gitlab.com/csltstu/sunine/-/tree/master/egs/voxceleb}.
Finally, we essentially obtain a 4-fold data size by combining the original clean data with three augmented copies.

\subsection{Batch sampler}

We then describe the batch sampler process for training. 
Each mini-batch $C$ is composed of randomly selected $S$ speakers, and from each speaker, $N$ utterances are sampled.
In our experiments, each speaker in $C$ only samples with one utterance ($N=1$).
Subsequently, DA with additive noise is applied to all the utterances in $C$, using a prior probability ratio of 6:4 between applying DA and not applying DA.
As depicted in the \emph{Batch sampler} module in Figure~\ref{fig:ada}, 
blocks of the same shape represent utterances from the same speaker, and blocks of the same color signify the same augmentation type.
Blocks with green color represent the raw clean data.

\subsection{Training Objective}

The training objective comprises two components: minimizing a speaker loss $L_{spk}$ and an adversarial augmentation loss $L_{adv}$.

For $L_{spk}$, both clean and augmented data from each speaker are passed through the embedding extractor to generate corresponding speaker embeddings.
These embeddings are then used to compute $L_{spk}$ by discriminating different speakers.
This process is essentially the same as the standard data augmentation training method.

Regarding  $L_{adv}$, its purpose is to remove the augmentation information from speaker embeddings.
Firstly, the augmentation classifier is employed to categorize different augmentation types.
Following that, a gradient reversal layer (GRL) is introduced between the embedding extractor and the augmentation classifier.
This GRL penalizes the ability of the augmentation classifier to correctly predict whether speaker embeddings come from the same augmentation types.
This ultimately results in the computation of the adversarial augmentation loss $L_{adv}$.

The overall loss is a linear combination of the speaker loss $L_{spk}$ and the adversarial augmentation loss $L_{adv}$ with a weight $\lambda$. 
In our experiments, $L_{spk}$ utilizes AAM-Softmax loss~\cite{xiang2019margin}, $L_{adv}$ is computed using binary cross-entropy, and $\lambda$ is set to 0.01.

\begin{equation}
L= L_{spk} + \lambda L_{adv}
\end{equation}

In summary, this training objective effectively combines standard data augmentation with adversarial training to 
allow the deep speaker model to learn speaker embeddings that are less sensitive to augmentation variations.
This enhances the robustness of deep speaker models, making them more resilient to the impact of acoustic variations.

\section{Experiments}

In this section, we present a comparison between the standard data augmentation (DA) and our proposed adversarial data augmentation (A-DA) 
on speaker verification under different test conditions. 

\subsection{Data}

\subsubsection{VoxCeleb~\cite{chung2018voxceleb2}}
This is a large-scale speaker dataset collected by the University of Oxford, UK. 
In our experiments, we used the development set of VoxCeleb2 (Vox2.dev) to train the x-vector models,
which includes a total of 5,994 speakers.
In addition, VoxCeleb1-O (Vox1.O) was employed as the validation trial set to select the optimal models, 
and VoxCeleb1-E/H (Vox1.E/H) served as the test trial sets to evaluate the model performance.

\subsubsection{CN-Celeb~\cite{fan2020cn}}
This is a multi-genre speaker dataset collected by Tsinghua University.
We used its standard evaluation set CNC.E, which consists of 200 speakers from 11 diverse genres, for performance evaluation.
Since the acoustic characteristics of CNC.E differ significantly from the VoxCeleb datasets, 
CNC.E can be used to validate the generalizability of the models.

\subsubsection{MUSAN~\cite{snyder2015musan} and THCHS30~\cite{wang2015thchs}}
The MUSAN database was used to sample interference signals for data augmentation, including three types: noise, music and speech.
Besides, we utilized the car and coffee noises from the THCHS30 database to create augmented test trials, which were used 
to assess the generalizability of the models.

\begin{table*}[htb!]
 \caption{EER(\%) results on VoxCeleb1 under different test conditions.}
 \label{tab:main}
 \centering
 \begin{tabular}{ll|ccccc|ccccc}
  \toprule[1pt]
  Method                &  Augtype  &  \multicolumn{5}{c|}{Vox1.E}              & \multicolumn{5}{c}{Vox1.H}               \\
  \midrule[1pt]
    -                   &  -        & Clean & Noise & Music & Speech & ALL     & Clean & Noise & Music & Speech  & ALL     \\
  \midrule[1pt]
  Baseline              &  -        & 1.363 & 2.876 & 1.843 & 2.423  & 2.213   & 2.300 & 4.866 & 3.322 & 4.131   & 3.778   \\
  \midrule[1pt]
  \multirow{3}{*}{DA}   & +noise    & 1.254 & 1.676 & 1.562 & 1.982  & 1.663   & 2.241 & 3.068 & 2.804 & 3.486   & 2.800   \\
								            & ++music   & 1.267 & 1.740 & 1.517 & 1.795  & 1.660   & 2.289 & 3.166 & 2.757 & 3.254   & 2.792   \\
								            & +++speech & 1.281 & 1.763 & 1.510 & 1.640  & 1.574   & 2.316 & 3.162 & 2.773 & 3.018   & 2.792   \\
  \midrule[1pt]
  \multirow{3}{*}{A-DA} & +noise    & 1.248 & 1.689 & 1.554 & 1.982  & 1.680   & 2.290 & 3.089 & 2.830 & 3.491   & 2.837   \\
								            & ++music   & 1.258 & 1.729 & 1.505 & 1.793  & 1.627   & 2.263 & 3.123 & 2.712 & 3.222   & 2.763   \\
								            & +++speech & 1.244 & 1.737 & 1.499 & 1.624  & 1.571   & 2.188 & 3.098 & 2.683 & 2.960   & 2.739   \\
         \bottomrule[1pt]
    \end{tabular}
\end{table*}

\subsection{Settings}

We followed the voxceleb/v2 recipe of the Sunine toolkit~\footnote{https://gitlab.com/csltstu/sunine/} to construct the speaker embedding model, 
which accepts 80-dimensional Fbanks as input features, adopts the ResNet34 topology for frame-level feature extraction, 
and uses the attentive statistics pooling (ASP)~\cite{okabe2018attentive} to produce speaker representations of x-vectors.
Once trained, the 256-dimensional activations of the last fully connected layer are read out as an x-vector. 
The simple cosine distance is used to score the trials in our experiments.

\subsection{Main results}

In our experiments, we constructed three groups of systems: one trained without DA (Baseline), 
one trained with standard data augmentation (DA), and one trained with our proposed adversarial data augmentation (A-DA).
For DA and A-DA, we gradually increased the diversity of augmentations to observe the performance trend.
The results in terms of equal error rate (EER) on the VoxCeleb evaluation datasets are reported in Table~\ref{tab:main}.

Firstly, it can be seen that both DA and A-DA methods significantly outperform the baseline, highlighting that 
the importance of data augmentation in enhancing model robustness. 

Furthermore, in nearly all test cases, A-DA consistently outperforms DA. 
More interestingly, this advantage becomes more pronounced with an increased diversity of augmentations.
This indicates that our proposed A-DA method can effectively mitigate the interference bias introduced by augmentation, 
thus further improving the robustness of the speaker embeddings.

\subsection{Further analysis}

To further validate the effectiveness of our proposed A-DA method, 
we conducted a series of performance evaluations under conditions of unseen augmentation types and more complex test conditions.
On one hand, we introduced additive noises using cafe and car sounds from the THCHS-30 dataset into the VoxCeleb test sets.
This was used to test the model's robustness to unseen augmentation variations.
On the other hand, we used the multi-genre CN-Celeb evaluation set, 
which has significantly different acoustics compared to the VoxCeleb training set 
and contains complex test conditions, such as multi-genre tests and cross-genre tests.
This aimed to assess the model's generalization performance.
The experimental results are reported in Table~\ref{tab:more} with EER as the performance metric.

\begin{table}[htb!]
 \caption{EER(\%) results under unseen augmentation variations and multi-genre test conditions}
 \label{tab:more}
 \centering
 \begin{tabular}{ll|cc|cc|c}
  \toprule[1pt]
  Method                &  Augtype   & \multicolumn{2}{c|}{Vox1.E} & \multicolumn{2}{c|}{Vox1.H} & CNC.E  \\
  \midrule[1pt]
  -                     &  -         & Car   & Cafe                & Car & Cafe                  &  -        \\
  \midrule[1pt]
  Baseline              &  -         & 1.423 & 2.274               & 2.518 & 3.897               & 13.461     \\
  \midrule[1pt]
  \multirow{3}{*}{DA}   & +noise     & 1.250 & 1.676               & 2.251 & 3.030               & 11.850     \\
                        & ++music    & 1.278 & 1.666               & 2.301 & 3.067               & 12.228     \\
                        & +++speech  & 1.286 & 1.667               & 2.340 & 3.095               & 12.397     \\
  \midrule[1pt]
  \multirow{3}{*}{A-DA} & +noise     & 1.260 & \textbf{1.672}               & 2.304 & 3.068               & \textbf{11.749}     \\
                        & ++music    & \textbf{1.270} & \textbf{1.660}               & \textbf{2.260} & \textbf{3.026}               & \textbf{11.963}     \\
                        & +++speech  & \textbf{1.254} & \textbf{1.639}               & \textbf{2.320} & \textbf{3.066}               & \textbf{12.154}     \\
  \bottomrule[1pt]
  \end{tabular}
\end{table}

Firstly, it can be seen that both DA and A-DA outperform the baseline under these more complex test conditions, 
providing further evidence for the effectiveness of the data augmentation technique.

Secondly, we can observe that for both unseen augmentation variations and the more complex multi-genre CNC.E test condition, 
A-DA still achieves a consistent performance advantage compared to DA, as indicated by the bold numbers. 
This demonstrates the strong robustness and generalization capability of A-DA compared with pure DA.

Finally, as the diversity of augmentation increased, DA and A-DA did not always achieve incremental improvements. 
This suggests that the learned speaker embeddings still contain some traces of augmentation variations, limiting their generalization to complex acoustic variations.
More appropriate training methods to address this augmentation residual issue should be explored for future research.

\section{Conclusion}

This paper introduces a new method that combines data augmentation with adversarial training, referred to as A-DA. 
It aims to address the issue of augmentation residual in vanilla data augmentation (DA), thereby improving robustness against complex acoustic variations. 
A-DA incorporates an augmentation classifier and utilizes a gradient reversal layer for adversarial training to decouple speaker information from augmentation variations, 
resulting in environment-invariant speaker embeddings. 
Experimental results demonstrated that A-DA outperforms DA in nearly all the test conditions, particularly in a more complex multi-genre condition represented by the CN-Celeb dataset, 
showcasing its robustness and generalization capability. 
Future work may involve augmenting with more complex acoustic variations (such as using AudioSet~\cite{gemmeke2017audio} for augmentation), 
and exploring techniques (e.g., mixed training~\cite{shi2023spot}) to further remove the acoustic variations from speaker embeddings.

\begin{acks}
This work was supported by the National Natural Science Foundation of China (NSFC) under Grants No.62171250 and No.62301075, and also the Huawei Cloud Research Program under project \\No.TC20220615035.
\end{acks}

\bibliographystyle{ACM-Reference-Format}


\end{document}